\documentclass[letterpaper,twocolumn,10pt]{article}
\usepackage{usenix-2020-09}
\usepackage{graphicx}
\usepackage{float}  
\usepackage{subfig} 
\usepackage{booktabs}
\usepackage{array}
\usepackage{makecell}
\usepackage{listings}
\usepackage{multirow}
\usepackage[frozencache,cachedir=minted-cache]{minted}
\usepackage{tikz}
\usepackage{amsmath}

\usepackage{pdfpages}

\usepackage{filecontents}

\begin{document}

\date{}

\title{\Large \bf BandwidthBreach: Unleashing Covert and Side Channels through Cache Bandwidth Exploitation}

\author{
{\rm Han Wang}\\
\and
{\rm Ming Tang\thanks{Corresponding author}}\\
\and
{\rm Ke Xu}\\
\and
{\rm Quancheng Wang}\\
\and
Key Laboratory of Aerospace Information Security and Trusted Computing, Ministry of Education, \\
School of Cyber Science and Engineering, Wuhan University
}

\maketitle

\begin{abstract}
In the modern CPU architecture, 
enhancements such as the Line Fill Buffer (LFB) 
and Super Queue (SQ), which are designed to track pending cache requests, 
have significantly boosted performance. To exploit this structure, 
we deliberately engineered blockages in the L2 to L1d route by controlling LFB conflict 
and triggering prefetch prediction failures, 
while consciously dismissing other plausible influencing factors. 
This approach was subsequently extended to the L3 to L2 and L2 to L1i pathways, 
resulting in three potent covert channels, termed L2CC, L3CC, and LiCC, 
with capacities of 10.02 Mbps, 10.37 Mbps, and 1.83 Mbps, respectively. 
Strikingly, the capacities of L2CC and L3CC surpass those of 
earlier non-shared-memory-based covert channels, 
reaching a level comparable to their shared memory-dependent equivalents. 
Leveraging this congestion further facilitated the extraction of 
key bits from RSA and EdDSA designs. 
Coupled with SpectreV1 and V2, 
our covert channels effectively evade the majority of traditional Spectre defenses. 
Their confluence with Branch Prediction (BP) 
Timing assaults additionally undercuts balanced branch protections, 
hence broadening their capability to infiltrate a wide range of cryptography libraries.

\end{abstract}

\section{Introduction}
\label{sec:intro}

Modern processors employ sophisticated optimization structures to enhance performance, 
introducing potential security vulnerabilities. 
Early research\cite{bernstein2005cache,percival2005cache,liu2015last,yarom2014flush+} 
concentrated on 
exploiting cache structures to launch attacks on cryptography libraries. 
More recent work\cite{paccagnella2021lord,gast2022squip,lipp2018meltdown,kocher2020spectre,bhattacharyya2019smotherspectre} 
has focused on compromising the isolation between processes. 
Notably, the Spectre \cite{kocher2020spectre} and Meltdown \cite{lipp2018meltdown} 
attacks demonstrated 
that even uncommitted secrets could be exposed. These groundbreaking discoveries have drawn 
considerable attention as they underscore a disconcerting possibility: 
flaws inherent in hardware 
could result in confidential information leaking from reliable to malicious processes, 
even in the absence of any software vulnerabilities.

A novel microarchitecture attack often relies on unexplored elements of the microarchitecture 
or leverages these elements in an entirely new way. 
Cache-based attacks fundamentally exploit the temporal difference between cached 
and non-cached data to discern secret data addresses. 
However, different attacks employ various methods to elicit this time disparity. 
In the Flush+Reload attack\cite{yarom2014flush+}, either the attacker or victim 
invalidates data in the cache through flushing or eviction, 
subsequently, the victim accesses the secrets, 
followed by the attacker trying to access the data using a known address. 
A cache hit signifies matching addresses, revealing the secrets' address to the attacker.
On the contrary, in the Prime+Probe attack\cite{liu2015last}, 
the attacker primes the cache with data at a known address, 
then the victim accesses the secret data, 
which may cause the eviction of data from the primed cache set. 
Afterwards, the attacker probes each cache set. If a cache miss is noticed, the attacker 
can infer the mapping of the secrets. 
Despite exploiting similar microarchitectural elements, and possibly the same feature, 
the distinct exploitation methods result in diverse attack scopes.

As attacks continue to rapidly advance, 
numerous countermeasures\cite{kiriansky2018dawg,taram2019context,saileshwar2019cleanupspec,barber2019specshield,fustos2019spectreguard,loughlin2021dolma} 
have been proposed to alleviate their impact. 
Cleaning the CPU microarchitectural state during context switches can thwart some attacks, 
as can cache partitioning and mapping randomization for specific cache-based attacks.
However, many attacks\cite{xu2022reverse,wang2023cross,gast2022squip,paccagnella2021lord} 
have exploited 
previously unexplored microarchitectural 
elements to bypass existing defenses. 
Lord of the Ring(s)\cite{paccagnella2021lord} was the first to reverse 
engineer the CPU's ring interconnect, 
discovering that data packets from different CPU cores could compete on the ring interconnect, resulting in time delays. 
This discovery facilitated the construction of a cross-core covert channel 
with a capacity exceeding 4 Mbps and a side-channel attack targeting cryptography libraries. 
Subsequent works \cite{wan2021volcano,wan2022meshup,dai2022don} have expanded upon this base, 
further reverse engineering and attacking server CPU mesh-connect structures. 
Another notable work\cite{xu2022reverse}, 
focuses on the contention issue in the transmission lines 
between a core's frontend and execution engine. 
These lines, dynamically shared between two logical cores, 
allow the sender and receiver to convey information 
through competition for the transmission lines.

The discussion of transmission lines doesn't stop here. 
Following the issuance of a load or store instruction, 
this instruction might sequentially access the L1, L2, and L3 caches, 
contingent on the state of storage. 
This pathway hasn't been seen as a bottleneck in execution time, 
because according to Intel manuals \cite{intel2023opt}, 
the actual transfer speeds between different cache levels 
fall below the designed maximum speed (64Byte/cycle). 
However, this viewpoint fails to consider the situations 
of Line Fill Buffer (LFB) saturation and hardware prefetching. 
LFB is a buffer utilized by the CPU to track and optimize pending memory requests, 
such as combining multiple stores or reads. 
We have reverse-engineered the specific number of LFB entries, 
with detailed discussion in Section \ref{sec:lfb-entries}. 
By creating contention for LFB and prefetching prediction failures, 
we successfully induced congestion in the cache bandwidth. 
Moreover, we've eliminated the influence of factors 
like execution port contention and L1 cache misses through experiments. 
Based on these findings, we extended our analysis, 
unearthing similar contention on the routes from L2 to L1 instruction cache (L1i) 
and from L3 to L2.

Our findings suggest that 
bandwidth contention leads to extended execution times 
for memory-related instructions in a process. 
Capitalizing on this observation, 
we encoded contention and non-contention states as binary '1' and '0' respectively, 
and created three distinct covert channels - L2CC, L3CC, and LiCC.
These channels respectively utilize bandwidth contention from L2 to L1 data cache (L1d), 
from L3 to L2, and from L2 to L1 instruction cache (L1i). 
Our covert channels deliver capacities up to 10.024 Mbps (L2CC), 10.369 Mbps (L3CC), 
and 1.838 Mbps (LiCC) from a single thread. 
To the best of our knowledge, both L2CC and L3CC outpace all previous covert channels 
that do not rely on shared memory\cite{xu2022reverse,paccagnella2021lord}, 
making them competitive with the fastest covert channels 
that do depend on shared memory\cite{saileshwar2021streamline}. 
We further demonstrate that our channels can maintain 
consistent transmission with minimal performance loss, even amidst substantial system noise.

Furthermore, cache bandwidth can also be exploited for side-channel attacks. 
We demonstrate the possibility of leaking key bits from RSA and EdDSA using cache bandwidth contention. 
When a cold start occurs or certain mitigation measures 
(such as mitigations to preemptive scheduling cache attacks) are enabled, 
the victim's instructions and data may miss in a specific cache level. 
This requires the victim to access a lower-level cache to retrieve subsequent instructions and data. 
At this point, an attacker can exploit bandwidth contention to leak the victim's key bits.

Aside from direct secret leakage in side-channel attacks, 
attackers often need to observe related data, 
such as timing or electromagnetic signals, 
and decode the secrets contained therein. 
In Spectre attacks, 
the victim's execution of secret-related code can cause out-of-bounds execution, 
leaving a trace in the cache. 
Without an observation mechanism, even if the victim performs an out-of-bounds execution, 
the attacker cannot obtain the secrets. 
Our covert channels can be combined with existing attacks 
to circumvent most cache-based Spectre defenses. 
We have demonstrated three potential new attacks as proof-of-concept (PoC). 
The first combines with SpectreV1, 
where the victim performs different memory-related operations 
during out-of-bounds and normal access. 
By monitoring their execution time, attackers can obtain the victim's secret. 
To reduce reliance on gadgets in the SpectreV1 attack, 
we combined our covert channels with SpectreV2. 
Attackers exploit branch target injection to force the victim to jump to 
a pre-selected gadget location, 
where the victim executes different memory-related operations based on the secret. 
The third attack manipulates the branch predictor to bypass balanced branches. 
When leaking key bits from RSA and EdDSA, 
the victim's code does not enable balanced branches as a defense against timing side-channels. 
We have demonstrated that, after enabling balanced branches, 
the attacker can discern whether the victim's 
speculative execution is successful, thereby obtaining the victim's secret.

This paper makes the following contributions:
\begin{itemize}
\item We present the first study focusing on cache bandwidth contention, 
and through reverse engineering, 
provide an extensive discussion on the causes of cache bandwidth congestion.
\item We have constructed three high-speed and noise-resistant covert channels, 
relying on cache bandwidth contention. 
L2CC and L3CC are, to the best of our knowledge, 
the fastest microarchitectural covert channels that do not depend on shared memory.
\item We leverage cache bandwidth contention to leak key bits from RSA and EdDSA. 
We have improved existing Spectre attacks and BP Timing attacks using our covert channels
to bypass several defenses and enhancing the attack potential.

\end{itemize}
The paper is organized as follows: 
Section \ref{sec:bg} introduces the relevant microarchitectural knowledge 
and related works. 
Section \ref{sec:reversing} discusses reverse engineering of the number of LFB entries 
and the causes of cache bandwidth congestion, 
and constructs three covert channels based on these findings. 
Section \ref{sec:exploit} details attacks on the RSA and EdDSA algorithms 
and proposes improvements to Spectre and BP Timing attacks. 
Finally, Section \ref{sec:conclusion} concludes and summarizes our research.

\section{Background}
\label{sec:bg}
\subsection{Introduction to microarchitecture}
\begin{figure}[htbp]  
    \centering  
    \includegraphics[width=.99\linewidth]{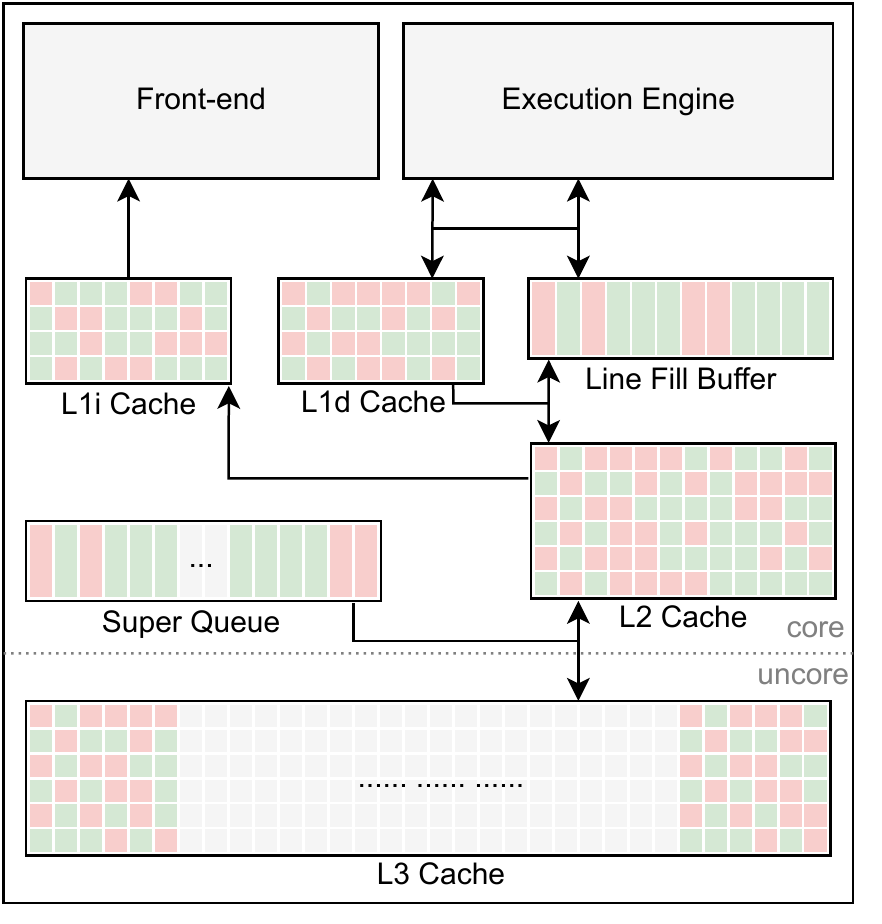}   
    \caption{Schematic diagram of the Intel x86 microarchitecture 
    focusing on the hierarchical construct of the memory subsystem.}  
    \label{fig:microarchitecture}
\end{figure}

\textbf{Cache} was introduced to boost processor performance and mitigate memory latency. 
As depicted in Figure \ref{fig:microarchitecture}, 
an x86 CPU's cache is structured into three levels - L1, L2, and L3 caches. 
Please note that the proportions in Figure \ref{fig:microarchitecture} 
do not imply any specific relationship regarding cache size. 
The L1 cache is divided into L1d and L1i caches, 
which store data and instructions respectively. 
However, this type-based differentiation is not present in L2 and L3 caches. 
L1 and L2 caches are located within the physical cores, while the L3 cache is outside the core, 
shared amongst multiple physical cores.

\textbf{Line Fill Buffer} (LFB) functions on the same level as the L1d cache. 
Missing read or write instructions in the L1d cache are stored here. 
If there are instructions targeting the same location, they can be merged. 
Lower-level cache hits promote cached data to the LFB for CPU utilization, 
followed by an L1d cache operation. 
The Super Queue (SQ) serves a similar function, 
except that its positioning is between the L2 and L3 caches.

\textbf{Micro-instruction Translation Engine} (MITE) 
is a part of the front-end decoding pipeline of a CPU core, 
translating complex x86 instructions into simple micro-ops. In Intel Skylake, 
there is one complex decoder and four simple decoders, 
each differing in the number of micro-ops they can issue per cycle. 

\textbf{Decoded Stream Buffer} (DSB) is an cache structure at the front end, 
which stores decoded instructions. When a hit is detected, MITE is disabled, 
allowing the DSB to directly transmit micro-ops to the Instruction Decode Queue(IDQ). 
Our LiCC works by increasing the number of instructions, effectively overwhelming the DSB, 
and forcing the front end to fetch instructions from the L1i and L2 caches for decoding.

Taking a memory access instruction as an example, 
the CPU front-end initially tries to retrieve micro-ops from the DSB. 
Failure to do so invokes retrieval from the L1i. In case of a miss in the L1i cache, 
queries are made to lower-level caches. 
When the instruction is executed in the back-end, 
queries are made to the L1d cache and the LFB simultaneously. 
When hitting, the data is immediately returned. 
If not, a slot is allocated in the LFB and a request is sent to access the L2 cache. 
A subsequent miss occupies a slot in the SQ, and the L3 cache is accessed.
Upon obtaining a hit, the data is progressively returned.

\textbf{Prefetching} is an optimization strategy within the CPU that boosts data access speed 
by predicting data usage patterns and proactively 
fetching data from lower-level into the higher-level cache. 
Prefetching can be divided into two types: software prefetching and hardware prefetching\cite{intel2023opt}. 
The hardware prefetching includes next-line prefetching, 
stride prefetching, adjacent-line prefetching, and stream prefetching. 
The first two types of prefetchers are situated in the L1d cache, 
and the last two types are found in the L2 cache. 
There are multiple methods to trigger hardware prefetching. 
In next-line prefetching, when an access to a particular line is detected, 
the subsequent line is prefetched. 
Stride prefetching sends prefetch requests to the next address based on the address, 
stride, and confidence level recorded in each entry. 
Adjacent-line prefetching prefetches the two lines neighboring the accessed line 
when an access to a particular line is detected. 
Stream prefetching conducts continuous 
prefetching according to the direction of the accesses.

\textbf{Cache Bandwidth} is divided into two types in the Intel manual\cite{intel2023opt}: 
peak and sustained bandwidth. 
For instance, the L2 cache has a peak bandwidth of 64B per cycle 
and a sustained bandwidth of roughly 29B per cycle. 
Peak bandwidth refers to the maximum data transfer capacity between the L2 and L1 caches per cycle, up to 64B. 
However, it doesn't imply that the 64B of data can be utilized by the core immediately. 
The sustained bandwidth, representing typical usage, 
indicates the time it takes for data to transition from L2 to L1d or the LFB 
and subsequently become available to the core.
The sustained bandwidth being less than 32B might be attributed to Intel 
considering the bandwidth loss caused by certain prefetchers. 
In our study, we consider the bandwidth congestion caused by LFB and SQ saturation, 
which may affect overall performance.

\textbf{Speculative execution} is a critical optimization measure 
employed in modern processors. 
Structures like branch prediction units enable processors 
to speculatively execute instructions out of order. 
The backend of the processor also contains a re-order buffer, 
which sequentially commits the state of each completed micro-op. 
Even though the outcomes of mispredicted executions are not committed, 
they may leave discernible traces within the processor's microarchitecture.

\textbf{Speculative execution attacks}, 
a notable exemplar being the Spectre-V1 attack\cite{kocher2020spectre}, 
transpire when an attacker manipulates the branch predictor, 
leading to incorrect execution of instructions by the victim. 
In such instances, despite the failure of array length checks, 
the CPU can be misled into speculatively executing out-of-bounds array accesses. 
Similar attacks are also witnessed in the context of indirect branches.

\subsection{Related Work}
\textbf{LFB-realted attacks.}
Previous LFB-related attacks were primarily focused on leaking information directly from 
the LFB. This flaw was initially exposed by RIDL\cite{van2019ridl}, 
in which the CPU speculatively loaded a value from memory, 
expecting it to be from a newly allocated page by the attacker, 
but it was actually from an LFB that belonged to a different process. 
ZombieLoad\cite{schwarz2019zombieload} improved upon RIDL using Meltdown. 
As prior LFB-related attacks couldn't leak data arbitrarily, 
CacheOut\cite{van2021cacheout} introduced a new cache pathway to ensure that attackers 
wouldn't have to wait for information in the LFB to become available, 
but could instead actively move it to the LFB using cache eviction. 
ÆPIC\cite{borrello2022aepic} identified Super Queue exhibiting undefined behavior 
while accessing bytes 4 to 15 of the APIC register, 
returning outdated data on Sunny Cove CPUs.
Our attack, unlike previous ones, 
doesn't directly leak data from LFB or Super Queue but utilizes contention to 
bypass potential defenses.

\textbf{Transmission-lines related attacks.}
Lord of the Ring(s) \cite{paccagnella2021lord} attack established a cross-core covert channel 
based on contention by reverse-engineering the protocol of ring interconnects. 
They also exploited the competition within ring interconnects 
to infer the victim's secret. Several subsequent 
papers \cite{wan2021volcano,wan2022meshup,dai2022don} have further examined 
and attacked the mesh structure in server CPUs.

Frontend Bus\cite{xu2022reverse} determined the shared strategy of the front-to-back 
transmission bus within a physical core by analyzing the 
effects of different loops on the execution time of the receiver. 
The authors found that the transmission bus from the front-end to the back-end 
within a physical core is dynamically shared between two logical cores. 
They proposed two new covert channels based on this discovery, 
but they didn't test the covert channels in a noisy environment.

Our covert channels also utilize transmission lines. 
However, we ensure superior transmission bandwidth 
and have conducted evaluations in noisy conditions.

\textbf{Prefetch-related attacks.}
Attacks targeting prefetchers are divided into those against hardware prefetching 
and software prefetching instructions. 
Hardware Prefetchers Leak\cite{bhattacharya2012hardware} revealed that prefetching 
initiates an inconsistency in the cache miss probabilities, 
leading to varied cryptographic operation execution time. 
Leaking Control Flow Information\cite{chen2021leaking} exploited Intel's instruction pointer-based stride prefetcher. 
A prefetcher trained in one domain can be used 
by another process in different security domains. 
A Fetching Tale\cite{cronin2019fetching} facilitated communication between a sender and receiver 
by training and evicting the trained prefetcher entries.

Prefetch Side-Channel Attacks\cite{gruss2016prefetch} exploited 
the execution time differences in prefetch instructions. 
These variances originate from that prefetch instruction needs to 
converts virtual addresses to physical ones. 
This resolving process sequentially 
scans several lookup tables and ceases once the address is located. 
Prefetching can thus determine if two virtual addresses 
correspond to the same physical address. 
AMD Prefetch Attacks\cite{lipp2022amd} 
adopted a similar approach but executed the assault on the AMD platform. 

Our exploit simply takes advantage of that 
prefetching will comsume the cache bandwidth, 
which remains uninhibited by existing prefetcher defenses.

\textbf{Cache defenses against speculative execution attacks.}
Several approaches are available to reinforce caches against Spectre attacks. 
DAWG\cite{kiriansky2018dawg} applies domain IDs to partition cache lines 
and ensure non-interference through cache replacement policies. 
CSF-LFENCE\cite{taram2019context} can inhibit speculative access 
by automatically inserting \textit{lfence}. 
For performance purposes, numerous hardware 
defenses\cite{fustos2019spectreguard,barber2019specshield} allow secret accesses 
but limit their subsequent execution usage or forwarding. 
DOLMA\cite{loughlin2021dolma} counters covert channels involving 
data caches, TLBs, instruction caches, 
and branch predictors. 
It mitigates alterations triggered by secret-dependent execution flows 
and resource contention, thereby effectively blocking cache-based side channels. 
CleanupSpec prevents speculative execution attacks from 
altering cache states by restoring the cache state when a speculation error is detected. 
In conclusion, based on the classification in \cite{hu2021sok}, 
defenses can be classified into 
No setup, No Access without Authorization, 
No Use without Authorization, and No Send without Authorization. 
These terms correspond to actions that obstruct covert channels or Spectre attacks setup, 
restrict speculative execution access to secrets, 
limit speculative execution use of secrets, 
and suppress transmission over covert channels, 
making it impossible for attackers to recover secrets.
The innovations presented in section \ref{sec:improve} demonstrate that 
combining with our covert channels, Spectre can surpass all defenses, 
excluding No Access without Authorization (CSF-LFENCE\cite{taram2019context}) 
and Delay in No Send without Authorization (DOLMA\cite{loughlin2021dolma}). 
These two categories have a huge performance overhead.

\section{Reverse Engineering and Covert Channels}
\label{sec:reversing}
\subsection{Reverse Engineering}
\label{sec:subreversing}
In this section, 
we will first determine the number of Line Fill Buffer(LFB) entries, 
which differs across various microarchitectures. 
Then, we will analyze the causes leading to cache bandwidth congestion, 
elucidating the role of the LFB in this context. 
\subsubsection{Determine the number of LFB entries}
\label{sec:lfb-entries}
The \textit{vmovntdq} instruction is a specific type of \textit{mov} instruction, 
which neither writes data to the cache nor retrieves 
the corresponding cache line from memory. 
Instead, it operates directly on memory. 
Utilizing the property that \textit{vmovntdq} tends to stay in the LFB 
for an extended duration, 
we can observe the contention of the LFB 
by controlling the number of \textit{vmovntdq} in each loop. 
The code is illustrated in Listing \ref{lst:vmov}. 

\begin{listing}[h]  
    \begin{minted}[linenos=false,frame=single]{nasm}
.loop:
    lea  rdi, [buffer]
    assign number 0
    rep repetitions
        vmovntdq [rdi+32+64*number], ymm0  
    assign number number+1
    endrep
    dec ecx
    jnz .loop
    \end{minted}
    \caption{Code to reverse engineer the number of LFB entries. 
    \textit{vmovntdq} resides in the LFB for a longer duration, 
    which allows us to observe the number of LFB entries by creating LFB saturation.}
    \label{lst:vmov}
\end{listing}

\begin{figure}[htbp]  
    \centering  
    \includegraphics[width=.99\linewidth]{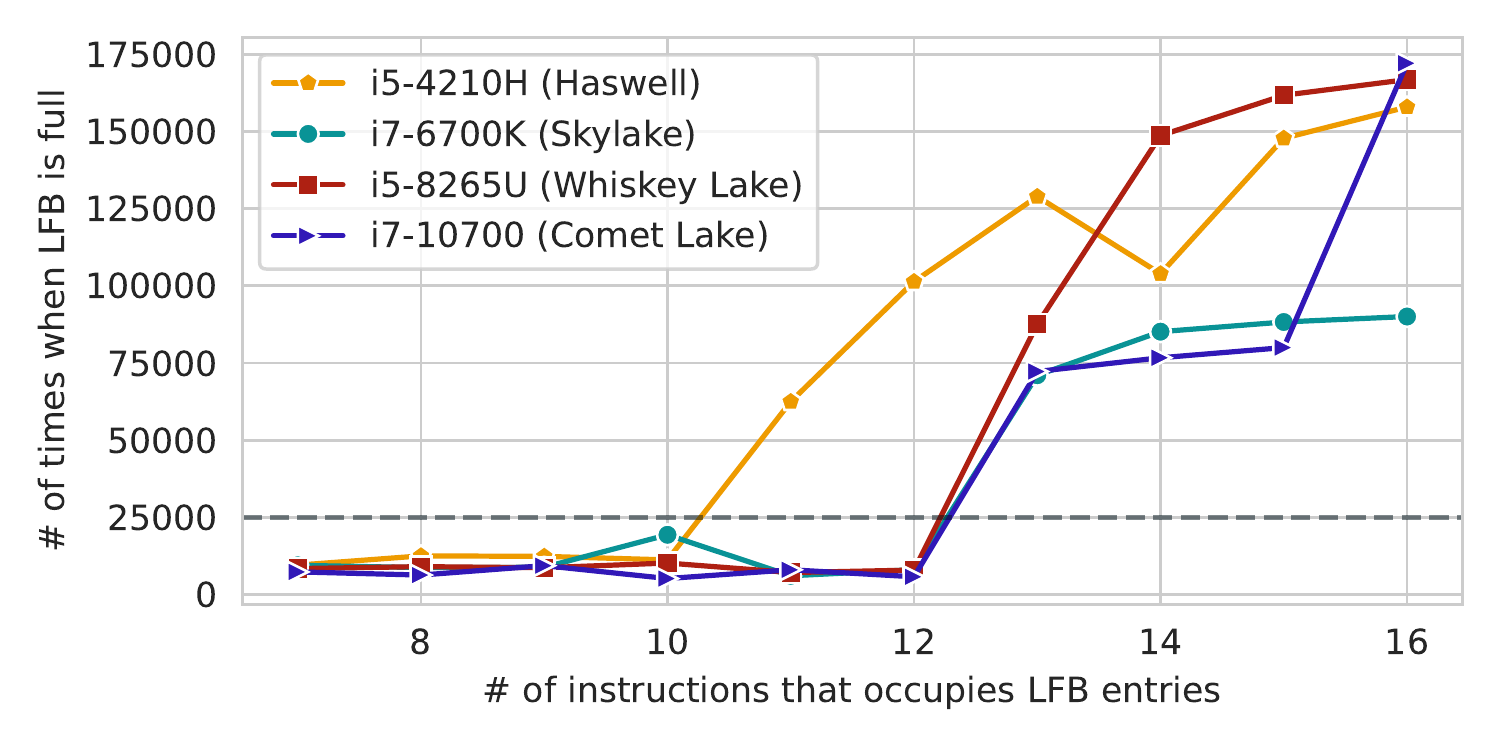}   
    \caption{Reverse-engineered results of the number of LFB entries. 
    A sudden increase in LFB saturation indicates that the LFB is full, 
    thereby providing the number of LFB entries.}  
    \label{fig:LFB-entry-result}
\end{figure}

When the number of \textit{vmovntdq} in a loop 
is fewer than the number of LFB entries, 
\textit{vmovntdq} will not be expelled from the LFB. As a result, 
an instruction writing at the same position in the next loop 
merges with the instruction in the LFB, 
attributed to the optimization feature of the LFB. 
When the number of \textit{vmovntdq} surpasses the number of LFB entries, 
earlier positioned instructions are expelled 
by later positioned ones within the same loop. 
Consequently, a \textit{vmovntdq} in the subsequent loop 
must reclaim an LFB entry, 
precipitating a substantial surge in the count when LFB is fully occupied.
The results for different microarchitectures are 
depicted in Figure \ref{fig:LFB-entry-result}. 
It can be observed that in the Haswell, 
a change occurs when the instruction count reaches 11, 
indicating that the number of LFB entries in the Haswell is 10. 
In contrast, starting from Skylake, 
Intel increased the number of LFB entries per core to 12.

\subsubsection{Analyzing the causes of bandwidth congestion}
\label{sec:cause}
Unless bearing an uncacheable tag, 
a read or write instruction may sequentially access the L1, L2, 
and L3 caches, contingent on the storage status. 
This data path has not been seen as a bottleneck 
in terms of execution time. 
Nonetheless, we've managed to induce congestion 
on this pathway and conducted experiments 
to discern the potential causes.  
All experimental data is exhibited in Table \ref{table:reverse1}, 
with all measurements taken on the receiver side. 
The sender and receiver are fixed on two logical cores 
of the same physical core.
\begin{table*}[t] 
  \newcommand{\tabincell}[2]{\begin{tabular}{@{}#1@{}}#2\end{tabular}}
  \centering 
  \caption{Reverse-engineered results from analyzing causes for contentions 
  in pathways from L2 to L1d and L3 to L2.  
  The difference between Group A and Group B is that the receiver 
  in Group A is allocated a buffer size of 128 KiB, 
  while in Group B it's allocated 512 KiB.}
  \label{table:reverse1}
  \begin{tabular}{llllllll}
      \toprule
      No & L1d Miss & L2 Miss & L3 Miss & LFB saturation & SQ saturation & Time & Description \\
      \midrule
      A1 & $14.8*10^9$ & $1.68*10^6$ & $53*10^3$ & $18.4*10^9$ & $N/A$ & $12.72s$ & \tabincell{l}{The sender and receiver access\\ L2 cache at the same time.} \\
      \specialrule{0em}{3pt}{3pt}
      A2 & $12.4*10^9$ & $0.08*10^6$ & $32*10^3$ & $9.8*10^9$ & $N/A$ & $4.73s$ & \tabincell{l}{The receiver accesses L2\\ cache alone.} \\
      \specialrule{0em}{3pt}{3pt}
      A3 & $14.3*10^9$ & $0.26*10^6$ & $52*10^3$ & $4.1*10^9$ & $N/A$ & $7.20s$ & \tabincell{l}{The sender accesses L1 cache while\\ the receiver accesses L2 cache.} \\
      \specialrule{0em}{3pt}{3pt}
      A4 & $12.4*10^9$ & $0.31*10^6$ & $230*10^3$ & $0.85*10^9$ & $N/A$ & $3.90s$ & \tabincell{l}{With prefetchers off, the sender\\ and receiver access L2 cache\\ at the same time.} \\
      \specialrule{0.001em}{3pt}{3pt}
      B1 & $15.0*10^9$ & $8*10^6$ & $139*10^3$ & $75.6*10^9$ & $1.4*10^9$ & $31.65s$ & \tabincell{l}{The sender and receiver access\\ L3 cache at the same time.} \\
      \specialrule{0em}{3pt}{3pt}
      B2 & $14.9*10^9$ & $6.7*10^6$ & $32*10^3$ & $49.2*10^9$ & $5.2*10^9$ & $11.79s$ & \tabincell{l}{The receiver accesses L3\\ cache alone.} \\
      \specialrule{0em}{3pt}{3pt}
      B3 & $14.8*10^9$ & $6.9*10^6$ & $53*10^3$ & $17.8*10^9$ & $2.7*10^9$ & $13.86s$ & \tabincell{l}{The sender accesses L1 cache while\\ the receiver accesses L3 cache.} \\
      \specialrule{0em}{3pt}{3pt}
      B4 & $15.0*10^9$ & $7.0*10^6$ & $230*10^3$ & $31.8*10^9$ & $14*10^3$ & $9.34s$ & \tabincell{l}{With prefetchers off, the sender\\ and receiver access L3 cache\\ at the same time.} \\
      \bottomrule
  \end{tabular}
\end{table*}

Based on our understanding of the microarchitecture, 
we hypothesize that when there's intense 
contention for LFB/SQ resources 
due to a large number of active memory accesses and hardware prefetching, 
the transmission path bridging higher and lower-level caches may experience congestion. 
\begin{table*}[htbp] 
    \newcommand{\tabincell}[2]{\begin{tabular}{@{}#1@{}}#2\end{tabular}}
    \centering 
    \caption{Reverse-engineered data analyzing the causes of congestion 
    in the L2-to-L1i path. 
    The CPU front-end is forced to fetch instructions from L1i and L2 
    as much as possible by increasing the number of instructions in the loop.}
    \label{table:reverse2}
    \begin{tabular}{llllll}
        \toprule
        No & L1i Miss & L2 Hit & Time & Cycles & Description \\
        \midrule
        C1 & $0.16*10^9$ & $7.2*10^3$ & $3.88s$ & $17.07*10^9$ & \tabincell{l}{The sender and receiver fetch from L2 cache\\ at the same time.} \\
        \specialrule{0em}{3pt}{3pt}
        C2 & $1.50*10^9$ & $2.3*10^3$ & $1.82s$ & $7.99*10^9$ & The receiver fetches from L2 cache alone. \\
        \specialrule{0em}{3pt}{3pt}
        C3 & $0.44*10^9$ & $14.7*10^3$  & $3.36s$ &$14.75*10^9$ & \tabincell{l}{The sender fetches from L1 cache while the \\receiver fetches from L2 cache.} \\
        \specialrule{0em}{3pt}{3pt}
        C4 & $0.14*10^9$ & $10.2*10^3$  & $4.27s$ &$18.80*10^9$ & \tabincell{l}{With prefetchers off, the sender and receiver\\ fetch from L2 cache at the same time.} \\
        \bottomrule
    \end{tabular}
\end{table*}

Experimentation will be employed to confirm this supposition. 
In A1, a 128KiB buffer is allocated to both parties, 
allowing the sender and receiver to simultaneously access the L2 cache. 
In A2, only the receiver operates independently, 
also with a 128KiB buffer allocated for accessing the L2 cache. 
The code, as shown in Listing \ref{lst:movzx}, 
uses larger stride lengths to mitigate the effects of L1 cache hits induced by prefetchers, 
aiming to maximize the number of memory access instructions that hit the L2 cache. 
At the same time, prefetching will also help us consume a portion of the bandwidth.

\begin{listing}[h]
    \begin{minted}[linenos=false,frame=single]{nasm}
%assign number 0
%rep repetitions
    movzx eax, BYTE [rdx+STRIDE*number]
    movzx ebx, BYTE [rdx+STRIDE_2*number]
%assign number number+1
%endrep
    \end{minted}
    \caption{Code for inducing L2-L1d and L3-L2 cache bandwidth congestion.}
    \label{lst:movzx}
\end{listing}

By comparing A1 and A2, 
we observed that the execution time of A1 is significantly higher than that of A2. 
However, this time increase could be attributed to multiple factors, 
including competition for front-end and back-end resources, 
a factor common to all hyper-threading timing side-channels, 
as well as L1d cache misses. It is necessary to rule out these factors.

In A3, the sender is allocated a 16KiB buffer, 
while the receiver retains a 128KiB buffer, 
allowing the sender to access the L1d cache and the receiver to access the L2 cache. 
Comparing A1 and A3, we found that A1's execution time remains higher than A3's, 
indicating that competition for front-end and back-end resources 
cannot sufficiently account for the difference in execution time.
There are other factors that contribute to the high execution time of A1.

In A4, the configurations for the sender and receiver are the same as in A1, 
but with the prefetchers disabled. 
We observe a significant decrease in execution time compared to A1. 
When comparing A1 and A4, 
we can identify two factors contributing to this reduction in time: 
fewer L1d misses and fewer instances of LFB saturation. 
However, when comparing A2 and A4, under the same number of L1d misses, 
A2 exhibits higher execution time even while having exclusive use of a whole physical core. 
This is due to the increased instances of LFB saturation in Group A2. 
Therefore, we conclude that prefetching and LFB saturation 
are the main causes of cache bandwidth congestion. 
This observation is more pronounced in Group B, 
where the impact of L1d misses appears to be even less significant.

Group B shares identical code and settings with Group A, 
with the sole difference being the size of the allocated buffers. 
Specifically, when accessing L1, a 16KiB buffer is allocated, 
while a 512KiB buffer is allocated when accessing the L3 cache. 

Similarly, the comparison between B1 and B3 indicates that the 
competition for front-end and back-end resources 
isn't the only source of the time differences; other factors, 
such as cache bandwidth competition, continue to impact the execution time in B1. 
In B4, the prefetchers are turned off, leading to a significant reduction in execution time 
compared to B1. 
However, none of the three cache levels' misses alone 
can account for the source of this time difference. 
The number of L1d misses is roughly equal, 
and the time difference caused by L2 and L3 misses 
is not on the same order of magnitude as the final time difference. 
On the other hand, the significant reduction 
in instances of LFB and SQ saturation can explain the decrease in execution time. 

\begin{figure*}[htbp]  
    \centering  
    \includegraphics[width=.99\linewidth]{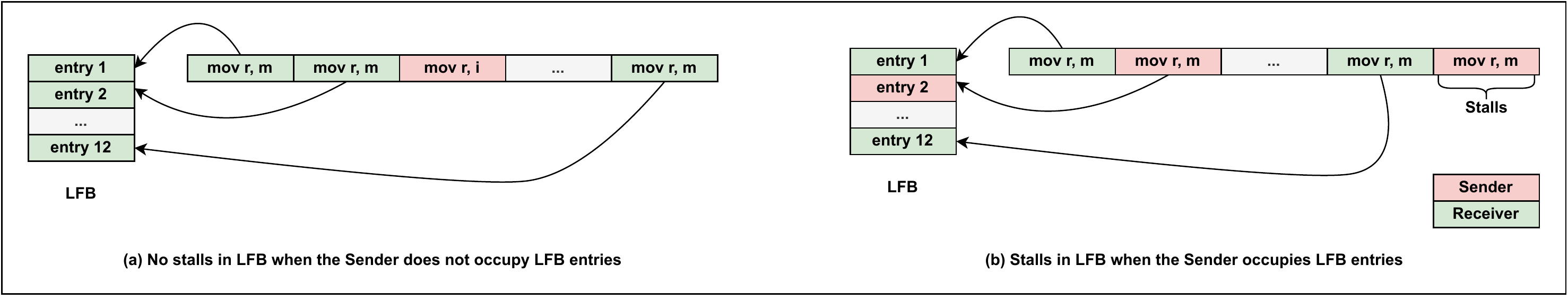}   
    \caption{Schematic diagram illustrating the principle of LFB contention.}  
    \label{fig:cache_contention_sr}
\end{figure*}

While there's no direct evidence, it's reasonable to suggest, 
based on the data from B4, that hardware prefetching might also occupy LFB and SQ entries. 
This doesn't mean that prefetching is the cause, 
and LFB/SQ saturation is the result; instead, 
LFB/SQ saturation exacerbates the time differences further. 
Therefore, we conclude that prefetching and 
LFB/SQ saturation are the primary causes of cache bandwidth congestion, 
which manifests as differences in the program's execution time.

\begin{listing}[h]
    \begin{minted}[linenos=false,frame=single]{nasm}
%assign number 0
%rep repetitions
    jmp .next_%+number
    nop11 
    nop11 
    nop11
    nop11 
    nop11 
    nop7
    .next_%+number :;
    %assign number number+1      
%endrep
.next_%+number :;
    \end{minted}
    \caption{Code for creating bandwidth congestion from L2 to L1i. 
    We have inserted long nop instructions to avoid backend execution 
    bottlenecks caused by jump instructions.}
    \label{lst:icache}
\end{listing}

Data from the L2 cache doesn't only feed the LFB and L1d caches, 
but also part of it enters the L1i cache. 
As a result, we analyzed whether the cache path from L2 to L1i 
could become congestion due to prefetching. 
Group C's configuration mirrors that of Groups A and B. 
By controlling the number of instructions to overwhelm the DSB, 
over 99.9\% of all instructions in Group C decoded in MITE pipeline. 
The code is shown in Listing \ref{lst:icache}.
This makes sure that, most of the time, 
the program is required to fetch instructions from caches other than the DSB. 
To avoid jump instructions from causing stalls in the backend, 
and therefore affecting time measurements, 
we inserted six long \textit{nop} instructions, totaling 62 bytes, 
between each \textit{jmp} instruction. 
Together with the \textit{jmp} instruction, this forms a 64B code block. 
Results are shown in Table \ref{table:reverse2}.

In C1, both the sender and receiver access the L2 cache simultaneously. 
A comparative analysis with C2, 
wherein only the receiver accesses the L2 cache, 
reveals an elevated execution time in C1. 
This suggests the existence of contention. 
However, the source of this contention is not yet clear, 
leading us to conduct experiments in C3. 
In C3, the sender attempts to fetch instructions from the L1 cache as much as possible, 
while the receiver aims to fetch instructions from the L2 cache. 
The potential causes of time discrepancies could be L2 cache hits and prefetching. 
However, considering that an L2 cache hit confers a mere 20-cycle advantage, 
which stands in stark contrast to the final cycle count, 
we posit that prefetching, which culminates in cache bandwidth congestion, 
serves as the principal influencer of the time differentials.
However, in the experiments conducted in C4, 
we did not observed the expected reduction in execution time after disabling the prefetcher. 
Despite ruling out the influence of caching factors, 
we posit that interference to the prefetcher occurs 
when simultaneously accessing the L2 cache, 
resulting in a decrease in bandwidth. 
This hypothesis is supported by a comparison of C1, C3, and C4. 
Similar observations were made during attacks on the cryptography libraries 
in Section \ref{sec:exploit}.

\subsection{Covert Channels}
\label{sec:cc}
\subsubsection{Threat Model}
\label{sec:threat-model}
\begin{figure}[htbp]
	\centering
	\subfloat[The transmission effect of L2CC.]{\includegraphics[width=.99\linewidth]{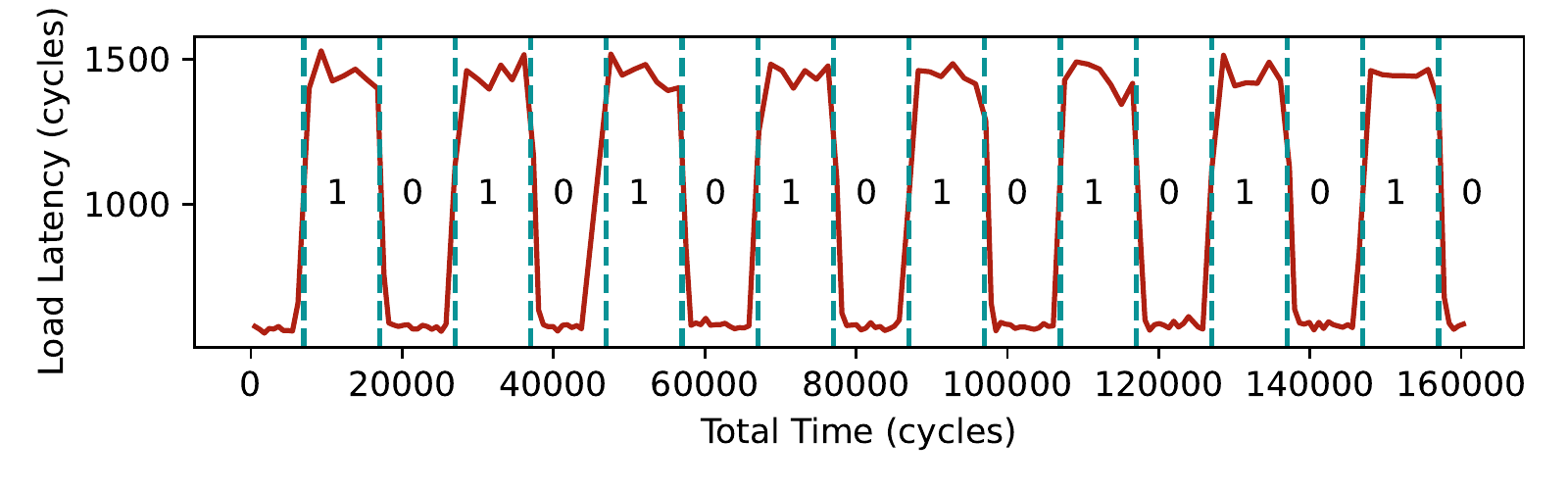}\label{fig:L2data-covert}}\\
	\subfloat[The transmission effect of L3CC.]{\includegraphics[width=.99\linewidth]{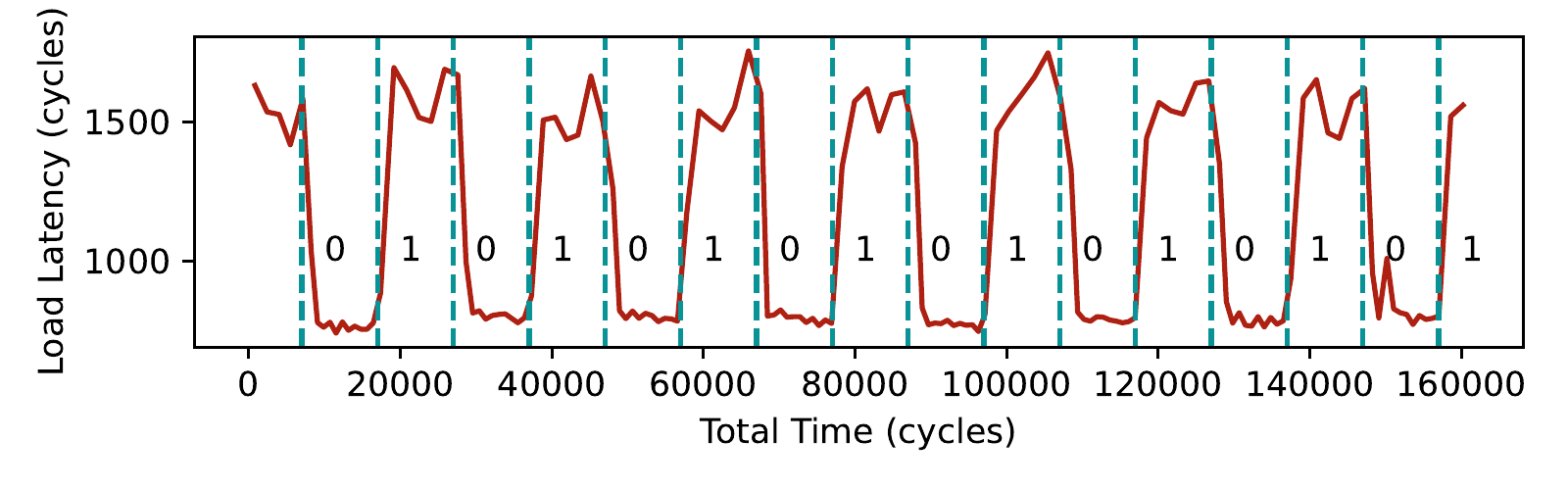}\label{fig:L3data-covert}}\\
	\subfloat[The transmission effect of LiCC.]{\includegraphics[width=.99\linewidth]{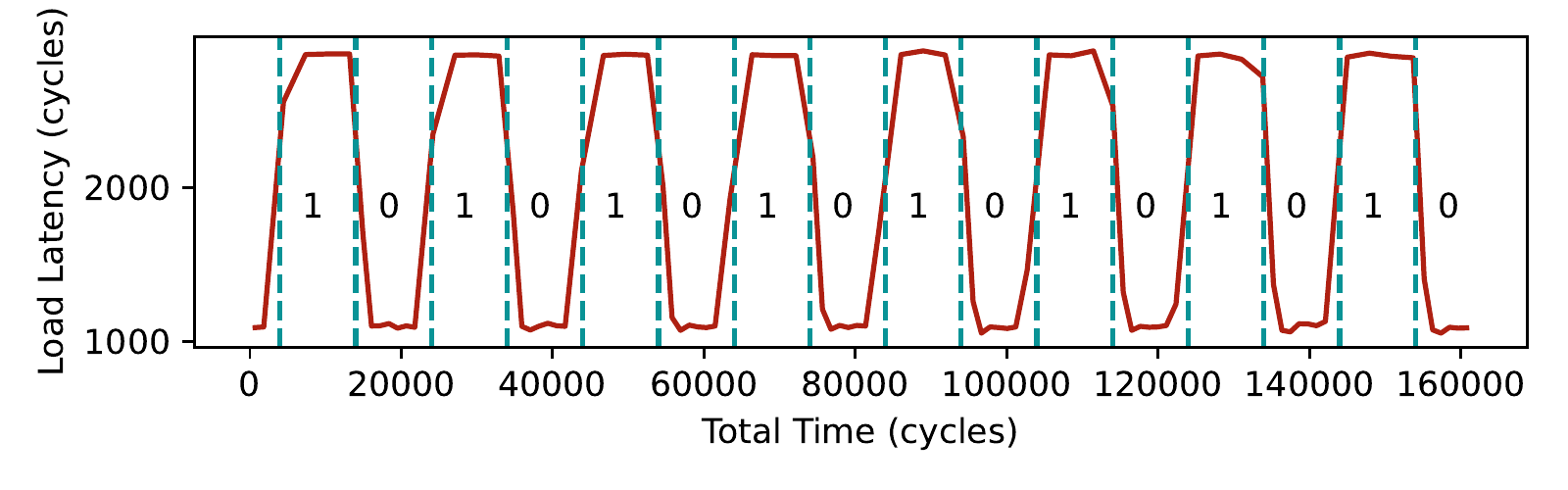}\label{fig:Lidata-covert}}\\
	\caption{ Transmission effects with a transmission interval of 10,000 cycles.}
    \label{fig:covert}
\end{figure}

\begin{figure*}[t]  
    \centering  
    \includegraphics[width=.99\linewidth]{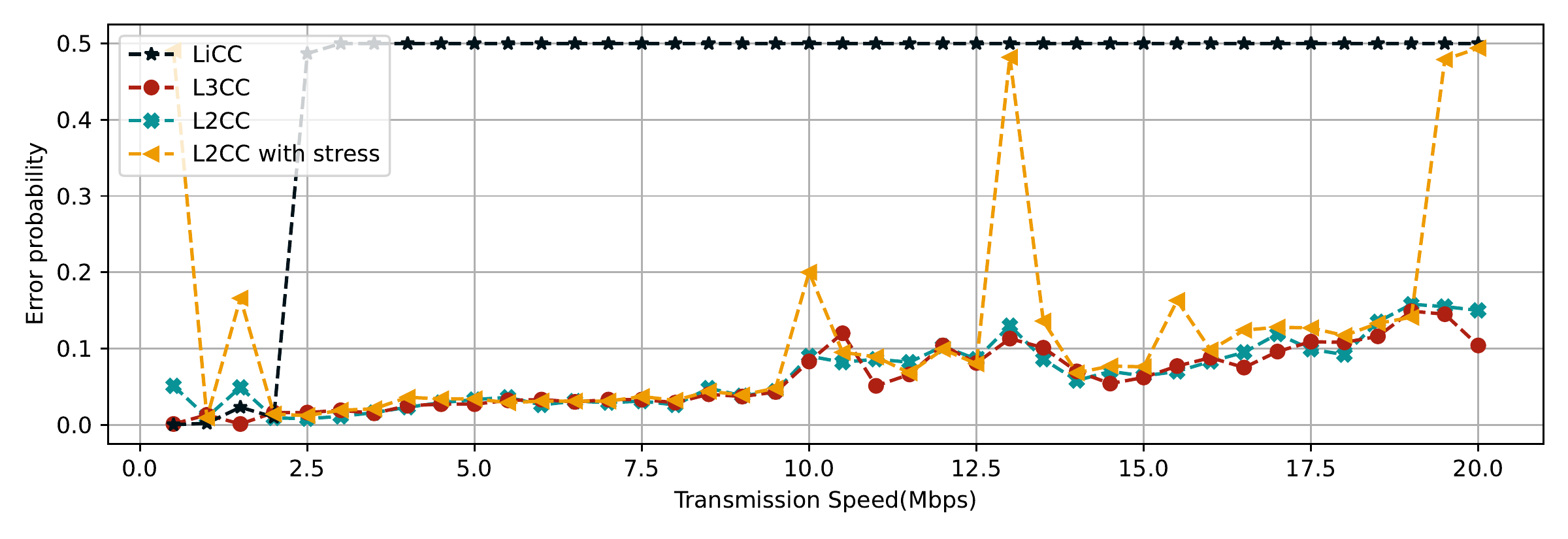}   
    \caption{The relationship between the transmission 
    speed and the error probability for our three covert channels. 
    We also measure the performance of L2CC in the presence of noise.}  
    \label{fig:error-plot}
\end{figure*}

In our three covert channels, 
we presume that the sender and receiver are located 
on different logical cores within the same physical core. 
We suppose both parties are considered to belong to different security domains 
and lack legitimate communication channels, 
in line with existing covert channel research\cite{xu2022reverse,paccagnella2021lord,wang2023cross,gast2022squip}. 
Our covert channel protocol resembles traditional cache-based covert channels, 
but uniquely, it does not necessitate shared memory between the sender and receiver, 
nor does it require root privileges.

\subsubsection{Configuration of Sender and Receiver}
\label{sec:configuration}
The sender and receiver convey information through contention on the cache bandwidth. 
We encode a '0' as a state with no contention on cache transmission 
and a '1' as a state with contention, illustrated in Figure \ref{fig:cache_contention_sr}. 
The receiver measures the time of memory access instructions, 
which consume a portion of the cache bandwidth. 
Consequently, when the sender transmits a '1', 
the memory access time for the receiver increases due to shared cache bandwidth usage. 
To synchronize both parties, we use a shared timestamp. 
There are other techniques \cite{hunger2015understanding,pessl2016drama} can replace this.

We first undertook a conceptual implementation of the covert channel 
to ensure accurate information transmission. 
In this scenario, the sender and receiver are threads 
running on two logical cores within the same physical core. 
In L2CC and L3CC, the sender and receiver engage in competition using the code from Listing 2, 
while both parties in LiCC utilize the code in Listing 3. 
The sender alternates between transmitting '1' and '0' every 10,000 cycles. 
In our three covert channels, '1' and '0' are distinguishable with clarity 
in Figure \ref{fig:covert}.

\subsubsection{Perfomance}
\label{sec:performance}
To assess the performance of our covert channels, 
we adopted a capacity metric, as referenced in \cite{paccagnella2021lord,okhravi2010design,pessl2016drama}, 
to measure the trade-off between the bit error rate and transmission performance. 
Capacity refers to the theoretical maximum throughput of a communication channel. 
We denote transmission speed as \textit{s}, error rate as \textit{e}, 
with capacity determined by the product of transmission speed and the binary entropy function.

\begin{align}
	&Capacity(s,e) = s \times (1-H(e)) \\
	&H(e) = -e \times \log_2(e) - (1-e)\times \log_2(1-e)
\end{align}

As depicted in the Figure \ref{fig:error-plot}, 
we present the performance of our three covert channels 
transmitting on i7-10700(Comet Lake, 4.00 GHz). 
An elevation in transmission speed incurs a gradual rise in the bit error rate. 
We have recorded both the maximum capacity 
and the capacity when the bit error rate is less than 5\% in Table \ref{table:compare}. 
Compared to related works\cite{xu2022reverse,paccagnella2021lord}, our capacity ranks the highest 
among covert channels that do not depend on shared memory, 
achieving up to 10.37 Mbps and 10.02 Mbps. 
In comparison to channels that utilize shared memory, 
we are on the same scale as the top-performing channel, Streamline\cite{saileshwar2021streamline}, 
without necessitating a large shared memory space of 64 MB as Streamline does.
\begin{table*}[t] 
    \newcommand{\tabincell}[2]{\begin{tabular}{@{}#1@{}}#2\end{tabular}}
    \centering 
    \caption{Comparison of covert channel capacities. 
    Our channel capacities significantly exceed those of prior channels 
    that didn't require shared memory, 
    and are on par with those channels that do require shared memory.}
    \label{table:compare}
    \begin{tabular}{llllll}
        \toprule
        Name & Bandwidth & Error & Capacity & Anti-noise & Shared Memory \\
        \midrule
        Lord of the Ring(s)\cite{paccagnella2021lord} & <6Mbps & <25\%  & 4.14Mbps & Not tested & No  \\
        \specialrule{0em}{3pt}{3pt}
        \multirow{1}{*}{Frontend Bus\cite{xu2022reverse}} & 0.7Mbps & 0.81\%  & 0.66Mbps & Not tested & No \\
                                     & 1.4Mbps & <10\%  & 0.75Mbps & Not tested & No \\
        \specialrule{0em}{3pt}{3pt}
        StreamLine\cite{saileshwar2021streamline} & 14.1Mbps & 0.37\%  & 13.57Mbps & Yes & Yes \\
        \specialrule{0em}{3pt}{3pt}
        Flush\&FLush\cite{gruss2016flush+} & 3.9Mbps & 0.84\%  & 3.60Mbps & Not tested & Yes \\
        \specialrule{0.00001em}{3pt}{3pt}
        \multirow{1}{*}{L2CC} & 18.0Mbps & 9.2\% & 10.02Mbps & Yes & No \\
                              & 9.5Mbps & 4.4\% & 7.03Mbps & Yes & No \\
        \specialrule{0em}{3pt}{3pt}
        \multirow{1}{*}{L3CC} & 20.0Mbps & 10.4\% & 10.37Mbps & Yes & No \\
                              & 9.5Mbps & 4.3\% & 7.07Mbps & Yes & No \\
        \specialrule{0em}{3pt}{3pt}
        LiCC & 2.0Mbps & 1.0\%  & 1.83Mbps & Not tested & No \\
        \specialrule{0em}{3pt}{3pt}
        \multirow{1}{*}{L2CC with stress} & 15.0Mbps & 7.6\% & 9.18Mbps & Yes & No \\
                                          & 9.0Mbps & 3.9\% & 6.86Mbps & Yes & No \\
        \bottomrule
    \end{tabular}
\end{table*}

Please note that real-world noise may reduce the channel's transmission speed 
and elevate the bit error rate. 
Furthermore, during the transmission with L2CC, 
we secured "stress -m" on the same logical core as the sender. 
The results, as shown in the Figure \ref{fig:error-plot}, 
indicate that our covert channel is mostly unaffected, 
showcasing its strong resilience to interference.

\section{Exploiting cache bandwidth contention}
\label{sec:exploit}
We propose two types of attacks, 
where the attacker's code bears similarity to the receiver outlined in Section \ref{sec:reversing}. 
The attacker consumes cache bandwidth via memory access instructions 
and measures their execution time. The attacker's buffer size is set to 128KiB. 
If the victim also executes memory access concurrently, 
the attacker's execution time will be extended. 
If the victim's memory access instructions are controlled by secrets, 
we can deduce the victim's secrets based on the execution time of the attacker. 
The experiments of Section \ref{sec:sca} were conducted on the i5-8265U, 
and those in Section \ref{sec:improve} on the i7-6700K. 
The experiments that combined with SpectreV2 required branch target injection. 
Although initial BTI attacks are mitigated in subsequent CPU generations, 
our work can be extended to newer CPUs via a fresh BTI attack vector\cite{barberis2022branch}.
Therefore, our Side channel can be applied to all CPUs possessing an LFB structure.

\subsection{Side channels}
\label{sec:sca}
\begin{figure*}[htbp]  
    \centering  
    \includegraphics[width=.99\linewidth]{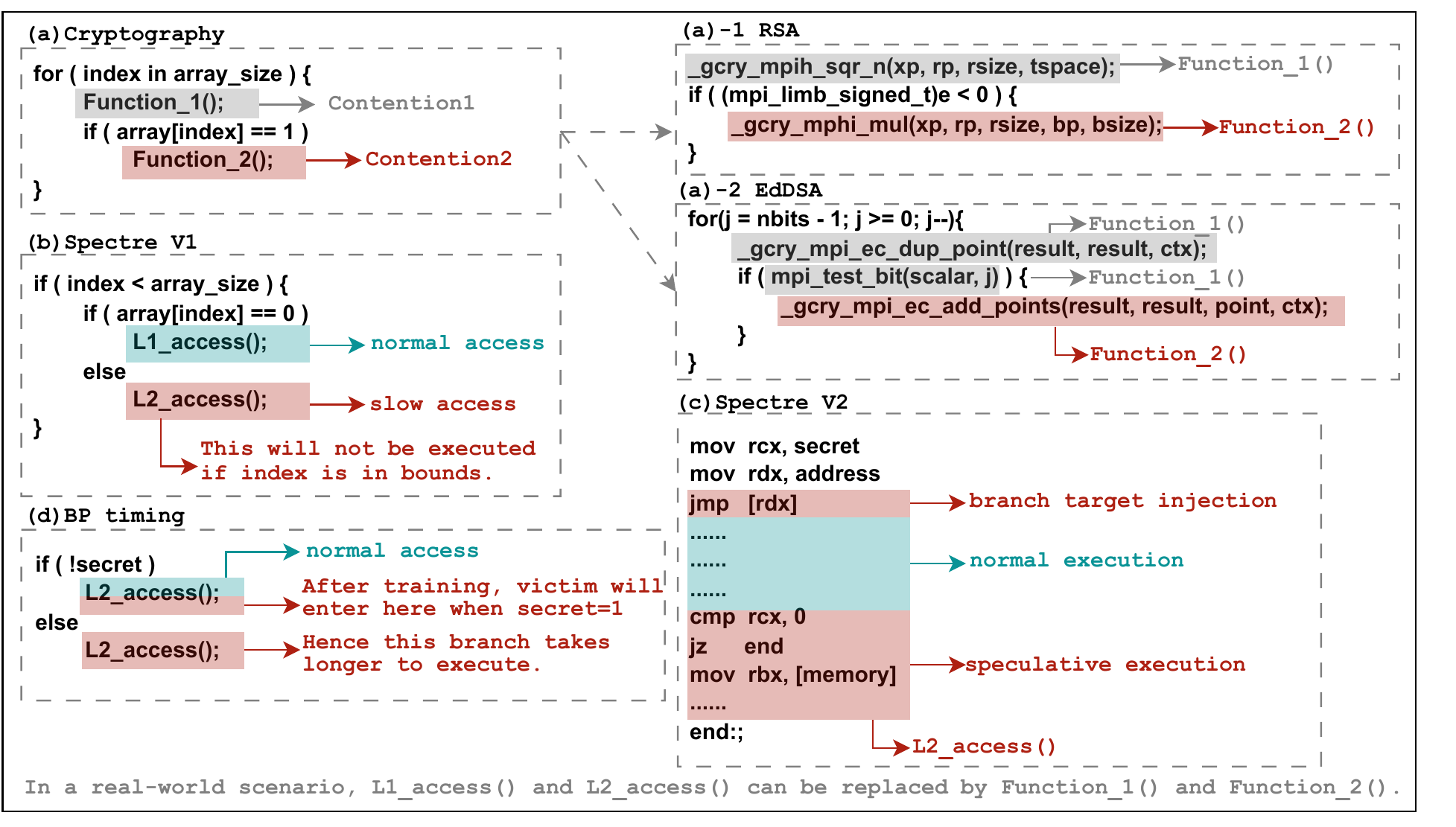}   
    \caption{Code patterns susceptible to exploitation through cache bandwidth contention.}  
    \label{fig:gadget}
\end{figure*}

First, we demonstrate that our side channel can effectively 
leak information from cryptography libraries. 
As shown in Figure \ref{fig:gadget}(a), 
many cryptography algorithms have adopted designs that follow such pseudocode, 
and there have been numerous side-channel attacks\cite{gras2020absynthe,gras2018translation,percival2005cache} 
previously launched against such implementations in cryptography libraries. 
Both \textit{Function\_1()} and \textit{Function\_2()} are 
related to cryptographic operations, while the array holds the key.

In a naive implementation, 
we first assume this is the first execution of the victim's code. 
The victim needs to fetch the instructions 
and relevant data from the off-core cache to the on-core cache, 
thereby occupying part of the cache bandwidth 
and consequently affecting the execution time of the attacker. 
The victim initially retrieves instructions and data related to \textit{Function\_1()}, 
instigating the first contention with the attacker. When the key is 0, 
the victim bypasses the execution of \textit{Function\_2()}, 
hence avoiding any cache bandwidth contention. 
When the key is 1, the victim needs to execute \textit{Function\_2()}, 
thereby fetching the relevant data and instructions, 
which leads to the second contention with the attacker. 
In the second cycle of the loop, since \textit{Function\_1()} has already been executed, 
no contention is observed when \textit{Function\_1()} is executed again. 
Hence, the attacker can infer whether the victim's key is 0 or 1 
based on the presence or absence of a second contention after the first.

As with many previous works \cite{evtyushkin2018branchscope,wang2019papp,paccagnella2021lord}, 
the attacker can utilize techniques such as preemptive scheduling 
to expand the leak to multiple key bits. 
By interrupting the victim to induce a context switch, the victim's cache is cleared, 
returning the cache state to that before the first loop.
Our implementation permits 
synchronization between the attacker and the victim. 
To achieve a cold start effect in experiments, the victim's cache is flushed before execution begins. 
We demonstrate the feasibility of using cache bandwidth contention 
for timing side-channel attacks on these two vulnerable versions 
of the RSA and EdDSA implementations.
The two implementations are illustrated in Figure \ref{fig:gadget}.

\begin{figure*}[ht]
	\centering
	\subfloat[Attack on the RSA algorithm leveraging L2-to-L1d bandwidth contention.]{\includegraphics[width=.32\linewidth]{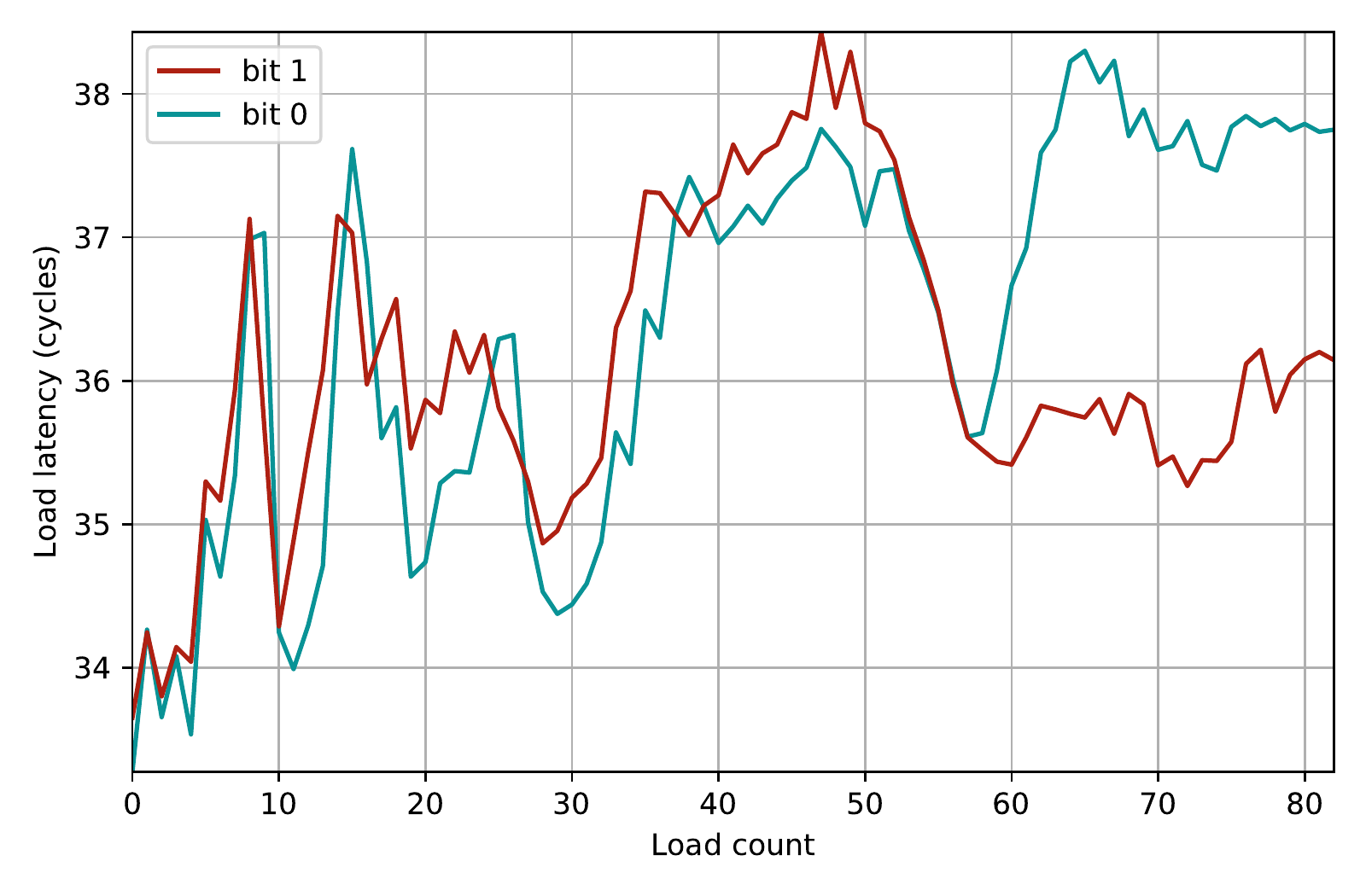}\label{fig:c1l2-128-count-1-rsa}}\hspace{1pt}
	\subfloat[Attack on the EdDSA algorithm using L2-to-L1d bandwidth contention.]{\includegraphics[width=.32\linewidth]{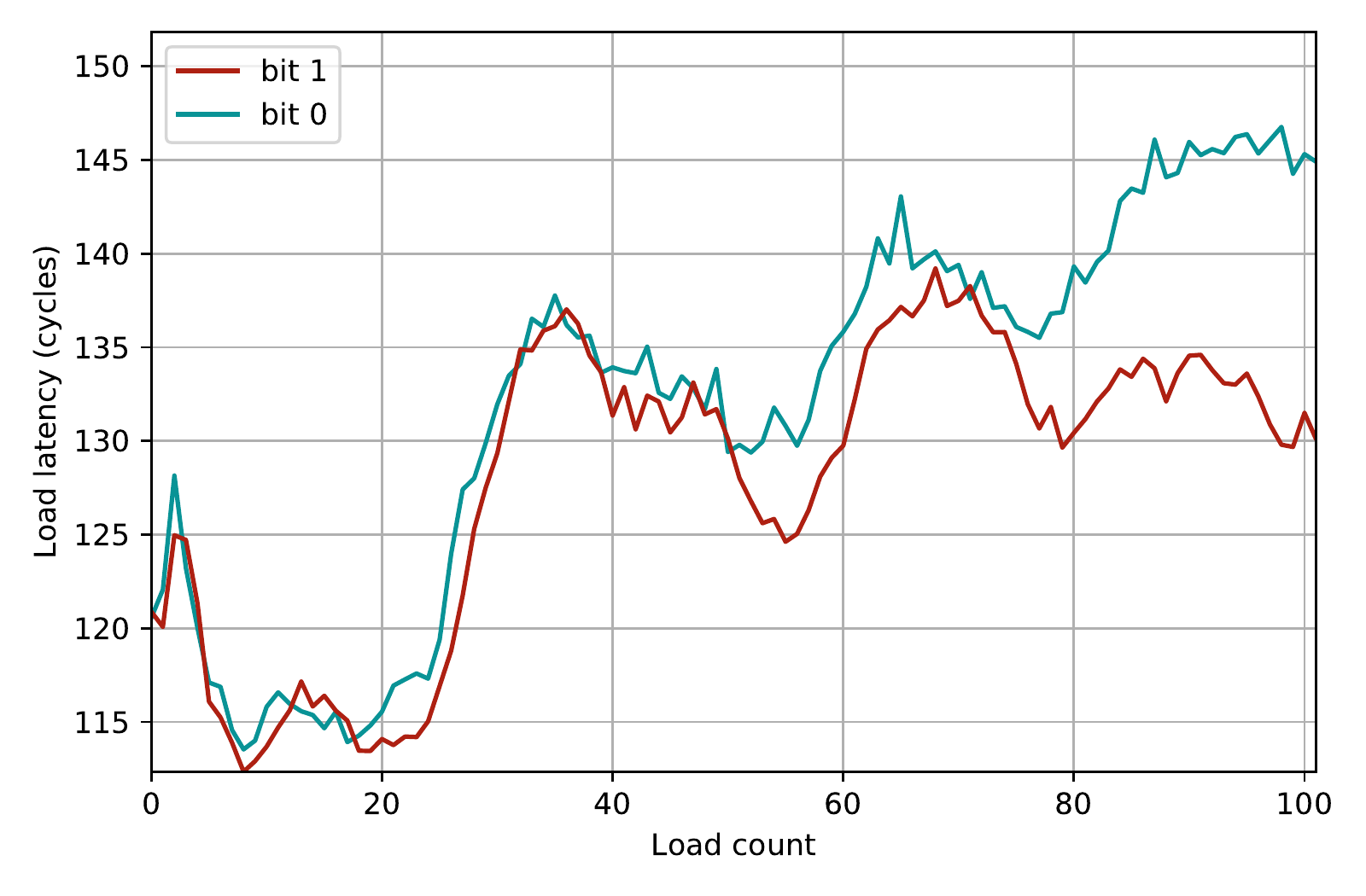}\label{fig:c1l2-128-eddsa-1}}\hspace{1pt}
	\subfloat[Attack on the EdDSA algorithm using L2-to-L1i bandwidth contention.]{\includegraphics[width=.32\linewidth]{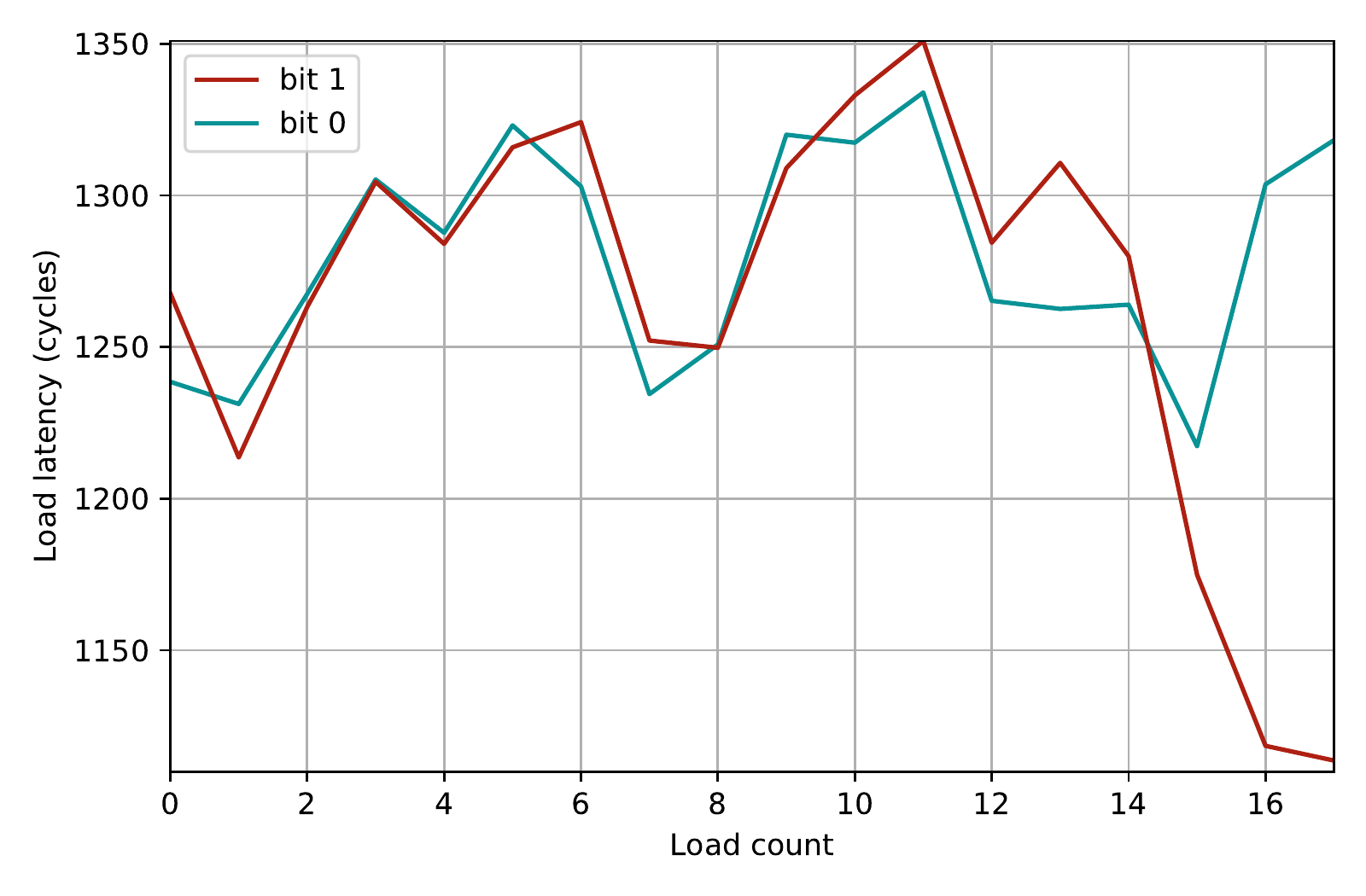}\label{fig:c1Licache-eddsa01}}\\
	\caption{Measured load latency during the victim's execution. 
    The displayed values represent the total latency of multiple load instructions 
    in the attacker's function.}
\end{figure*}

Figure \ref{fig:c1l2-128-count-1-rsa} exhibits the result of the attack 
on the RSA in libgcrypt1.5.2 utilizing contention over L2 to L1 cache bandwidth. 
The attacker's timing function is the same as in Listing \ref{lst:movzx}. 
The repetition is set to 8, 
with a loop count of 1. 
Up to the 30th call, the timing is unstable, 
as we're still in the phase before entering the target gadget. 
From the 30th to the 60th calls, 
the contention between the attacker and the victim's \textit{\_gcry\_mpih\_sqr\_n()} is documented. 
A discernible peak is observable due to the victim's 
unconditional execution of squaring. 
Intriguingly, after the 60th call, 
we witness a situation completely opposite to our expectations, 
which can be reproduced consistently. 
When the bit is 1 and executing \textit{\_gcry\_mphi\_mul()} requires 
fetching instructions and data into the on-core caches, 
the execution time is unexpectedly shorter than when no fetching occurs. 
We theorize that this may be because the attacker's prefetching 
remains undisturbed when \textit{\_gcry\_mpih\_sqr\_n()} 
is executed twice consecutively (bit=0), 
whereas the prefetching is somewhat disrupted 
when \textit{\_gcry\_mpih\_sqr\_n()} and \textit{\_gcry\_mphi\_mul()} 
are run in sequence, causing a reduction in cache bandwidth contention. 
As this counterintuitive peak can occur consistently, 
it does not hamper the leakage of secret information. 
For the RSA, we gathered 3000 traces, split them into training and testing sets, 
and trained a support vector machine. 
On the test set, the L2 cache bandwidth contention 
yields an accuracy rate exceeding 97.04\%.

We also targeted the EdDSA Curve25519 in libgcrypt 1.6.3. 
The function \textit{\_gcry\_mpi\_ec\_mul\_point()} contains a secret-dependent branch. 
Its structure is generally akin to the one depicted in our Figure \ref{fig:gadget}(a)-2. 
There are two \textit{Function\_1()} in it.
\textit{gcry\_mpi\_ec\_dup\_point}is executed once, followed by \textit{mpi\_test\_bit()}. 
The addition function \textit{gcry\_mpi\_ec\_add\_points()} 
is conditionally executed only when the key bit is 1. 
As evident in Figure \ref{fig:c1l2-128-eddsa-1}, 
two distinct peaks emerge when the key bit is 1, 
corresponding to the two executions of \textit{Function\_1()}. 
Similar to the attack on RSA, 
the third peak also presents a longer execution time when bit=0. 
We speculate that this might be because the attacker's 
prefetching is somewhat disrupted when \textit{Function\_1()} and \textit{Function\_2()} 
are executed consecutively. 
Given the lengthy execution time of this function, 
we also utilized L1i cache bandwidth contention for the attack. 
As shown in Figure \ref{fig:c1Licache-eddsa01}, the main distinction in the waveform emerges after the 14th cycle, 
where an obvious upward peak can be observed when the key is 0. 
For EdDSA, we adopted the same testing and training method as for RSA. 
Contention on the L2 cache bandwidth yielded an accuracy rate of 91.60\% on the test set, 
while contention on the Li cache bandwidth resulted in a 75.9\% accuracy rate. 
The lower accuracy rate could potentially 
be due to the lack of sufficient distinction in the absence of LFB and SQ structures.

\subsection{Improvments to existing attacks}
\label{sec:improve}
In this section, we improve SpectreV1, SpectreV2, 
and Branch predictor Timing attacks in an attempt to bypass existing defenses.

A side-channel attack comprises three steps: 
the attacker set up, the victim executes, 
and the attacker observes the corresponding data and decodes the secret from it. 
Using the classic SpectreV1 implementation as an example, 
the attacker first trains the branch predictor by repeatedly executing the branch. 
The victim's execution leads to out-of-bounds memory accesses, 
which leaves a trace in the cache. 
The attacker then needs to observe the cache to obtain the secret. 
That is to say, without an observation method, 
even if the victim performs an out-of-bounds memory access, 
the attacker cannot obtain the secret.
\begin{figure}[!htbp]  
  \centering  
  \subfloat[Distribution plot of L2CC combined with SpectreV1.]{\includegraphics[width=.99\linewidth]{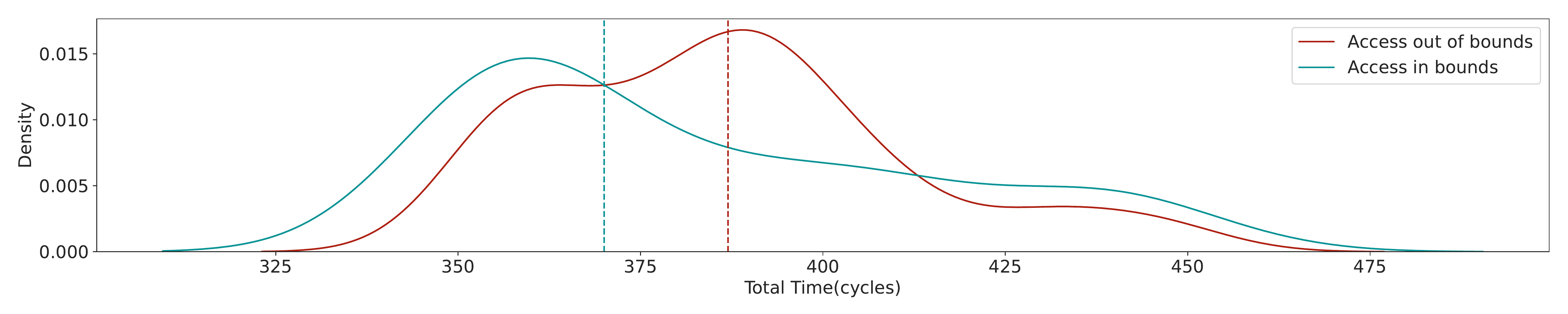}\label{fig:final-spectre-v1-contention}}\\
  \subfloat[Distribution plot of L2CC combined with SpectreV2.]{\includegraphics[width=.99\linewidth]{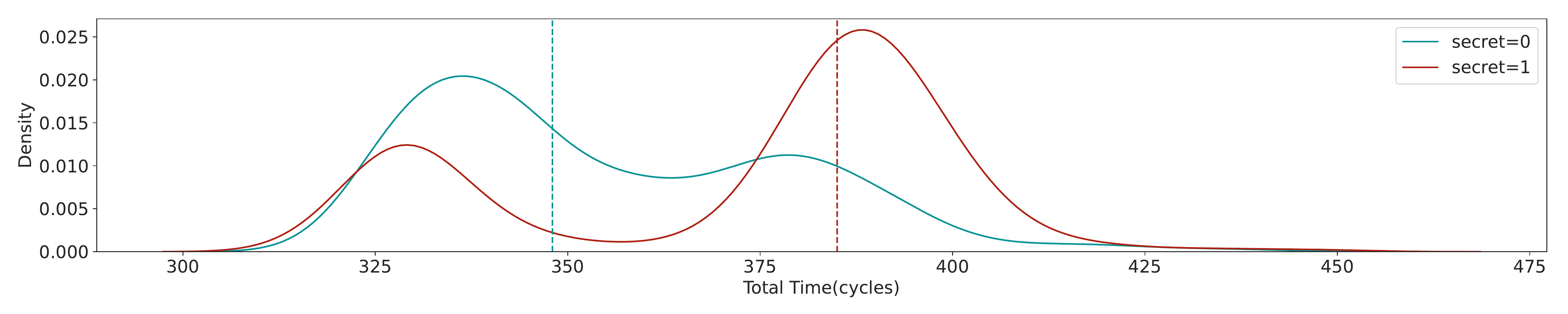}\label{fig:final-spectre-v2-contention}}\\
  \subfloat[Distribution plot of L2CC combined with BP Timing attack.]{\includegraphics[width=.99\linewidth]{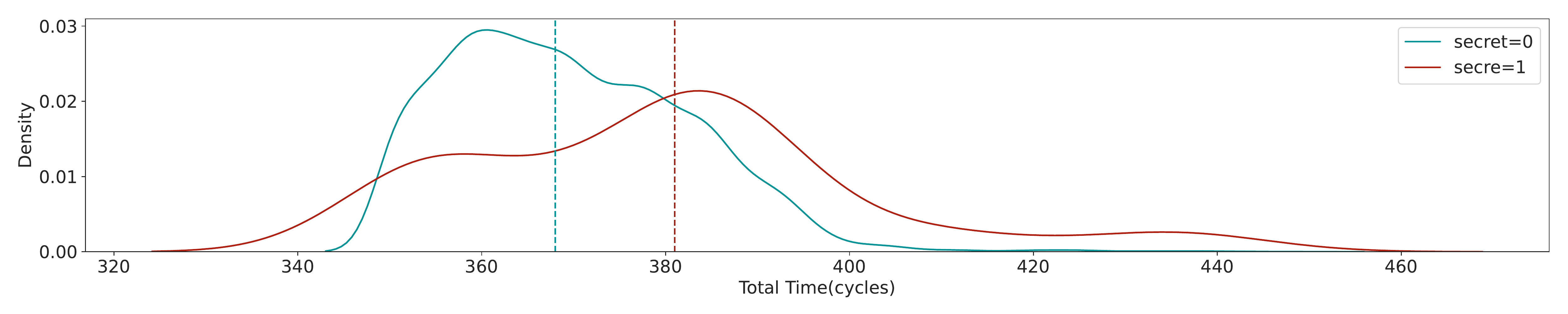}\label{fig:final-spectre-v1-contention-new}}\\
  \caption{Results of improvements to existing attacks. 
  Distribution plots obtained from multiple executions demonstrate the potential 
  for differentiating victim secrets.}
  \label{fig:3spectre}
\end{figure}

\cite{hu2021sok} outlines four distinctive categories of defense: 
No setup, No Access without Authorization, 
No Use without Authorization, and No Send without Authorization. 
These are explained in Section \ref{sec:bg}. 
Except for defenses in the branch predictor, 
the remaining defenses aim to disrupt the covert channels in Spectre. 
Our improvements to Spectre attacks enable us to 
bypass all these defense mechanisms except for the No Use without Authorization 
and Delay in No Send without Authorization, 
which have a huge performance overhead.

We propose three new combinations to bypass the shielding of observation methods. 
The first one involves a modification to SpectreV1, 
which allows the attacker to avoid sharing memory with the victim. 
Even if the victim possesses a separate memory partition, 
as seen in some defenses, the secret can still be leaked. 
The code is illustrated in Figure \ref{fig:gadget}(b). 
In this implementation, during normal execution 
the victim continuously executes the \textit{L1\_access()} function 
without occupying the attacker's cache bandwidth. 
However, once training is completed 
and out-of-bounds execution occurs to execute \textit{L2\_access()}, 
the attacker's cache bandwidth becomes occupied, 
extending the execution time and thus revealing the secret.

Despite the adaptations to SpectreV1 requiring certain gadgets in the victim's code, 
this restriction is relaxed when our covert channel is combined with SpectreV2. 
As illustrated in Figure \ref{fig:gadget}(c), 
the victim initially moves the secret into the rcx register and 
jumping to the address stored in rdx. 
However, due to the influence of the attacker, 
the control flow is redirected to a gadget in the victim's code. 
The speculative execution causes cache access, affecting the attacker's cache bandwidth. 
Therefore, by observing the execution time, the attacker can deduce the victim's secret.
In comparison to the SmoTherSpectre\cite{bhattacharyya2019smotherspectre}, 
which utilizes port contention for transmission, 
our approach exhibits higher resilience against system noise and interference 
from other processes.

For branches that are dependent on secrets, 
a common code-level defense is branch balancing. 
This is demonstrated in Figure \ref{fig:gadget}(d), 
which depicts a typical balanced branch structure. 
This defense strategy can be easily integrated into the cryptography libraries 
described in Section \ref{sec:sca}. 
Nevertheless, we have managed to overcome this defense by training the branch predictor. 
When the secret is 0, the victim carries out regular L2 access. 
Yet, after the attacker trains the branch predictor, 
if the secret is 1, due to speculative execution, the victim enters the 'if' branch, 
competing with the attacker for cache bandwidth. 
The victim then re-executes the 'else' branch. 
As a result, the execution time for the attacker is longer 
than when the victim only executes the 'if' branch, 
thereby making it possible to distinguish the secret.

The results are depicted in the figure \ref{fig:3spectre}, 
showing distinct time distributions for the attacker in both scenarios, 
hence allowing differentiation. 
It's important to highlight that both L1 and L2-access functions 
can be replaced with specific function implementations, 
which would further accentuate the time differences.

\section{Conclusion}
\label{sec:conclusion}
In this study, we analyze the root causes of cache congestion, 
establishing cache path congestion utilizing LFB/SQ contention 
and prefetch prediction failures. Such contention enabled us 
to construct three high-speed, noise-resistant covert channels. 
Our capacity ranks the highest among covert channels that do not depend on shared memory, 
yielding up to 10.37 Mbps and 10.02 Mbps. 
When measured against channels reliant on shared memory, 
our channels align closely with the performance of the highest-ranking channel, 
Streamline\cite{saileshwar2021streamline}, 
without necessitating a large shared memory space of 64 MB as Streamline does. 
Additionally, we employed the contention to extract key bits from RSA 
and EdDSA implementations. Finally, we integrated our approach 
with existing Spectre and BP Timing attacks, circumventing some existing defenses.

A fundamental strategy for defense generally entails the avoidance of 
supplying high-precision clocks. However, given the high bandwidth of our covert channels, 
it's still possible to discern differences on low-precision clocks by increasing 
the number of iterations.
In terms of potential defensive strategies, 
the most direct approach would be to disable hyper-threading. 
However, this method comes with a significant performance overhead. 
As an alternative, one could consider implementing static partitioning 
of the Line Fill Buffer (LFB) and Super Queue (SQ). 
This could help prevent cache bandwidth contention that may arise 
when attackers occupy the entire LFB/SQ structure. We have disclosed our results to Intel.





\bibliographystyle{plain}
\bibliography{\jobname}

\end{document}